\documentclass[12pt]{article}
\usepackage{graphics}
\usepackage{epsfig}
\usepackage{epstopdf}
\usepackage{amsmath}
\usepackage{array}
\usepackage[numbers,sort&compress]{natbib}
\begin{document}
\begin{center}
{\bf \Large {Bulk viscous matter and recent acceleration of the universe based on causal viscous theory}} \\[0.2in]
Jerin Mohan N D, Athira Sasidharan and Titus K Mathew \\
Department of Physics, Cochin University of Science and Technology, Kochi-22, India.\\
{\it jerinmohandk@cusat.ac.in, athirasnair91@cusat.ac.in, titus@cusat.ac.in} \\[0.3in]
\begin{abstract}
Evolution of the bulk viscous matter dominated universe has been analysed using the full causal, Israel-Stewart theory for the evolution of bulk viscous pressure in the context of recent acceleration of the universe. The form of bulk viscosity is taken as $\xi=\alpha\rho^{1/2}. $ We obtained analytical solutions for the Hubble parameter and scale factor of the universe. The  model parameters have been computed using the Type Ia supernovae observational data. The evolution of the prominent cosmological parameters were obtained. The age of the universe for the best estimated model parameters is found to be less than observational value. The viscous matter behaves like stiff fluid in the early evolutionary phase and then evolves to a negative pressure fluid in the later phase. The equation of state is found to be stabilized with value $\omega >-1$ and thus, the model is not showing any of the phantom behaviour during the evolution. The local as well as generalized second law of thermodynamics are satisfied in this model, while it was shown by many that the local second law is breaking in the Eckart formalism approach. The statefinder geometric diagnostic shows that the present model is distinct from the standard $\Lambda$CDM model of the universe. One of the marked deviation seen in this model compared to a corresponding model using Eckart approach is that the bulk viscosity decreases with expansion of the universe, while in it increases from negative value in the early universe towards positive values values in the Eckart formalism.
\end{abstract}
\end{center}
\section{Introduction}
The observational data indicates that the present universe is expanding as well as accelerating\cite{{Reiss},{Perlmutter}}. Many theoretical models have been proposed to interpret this recent acceleration either by modifying the right hand side of the Einstein's gravity equation with specific forms of the energy momentum tensor $T_{\mu\nu}$ which can cause a negative pressure or by modifying the left hand side i.e. the geometry of the space time. In the first approach, one need an exotic cosmic component, dubbed as ``dark energy", with equation of state satisfying,
$\omega<-\frac{1}{3}.$ The most successful model of the universe, which explains the recent acceleration of the universe is the standard $\Lambda$CDM model. This model incorporates the cosmological constant $ \Lambda $, characterized by the equation of state $ \omega_{\Lambda}=-1,$ as the dark energy. Even though this model fitting substantially well with the 
observational data, it is faced with some drawbacks, mainly the coincidence problem and the cosmological constant problem. The model is unable to explain the observed coincidence between the densities of non-relativistic matter and cosmological constant of the current universe, known as the coincidence problem. The cosmological constant problem is regarding the large discrepancy between the theoretically predicted value of the cosmological constant and it's observed value. The value of the cosmological constant predicted from field theoretical estimation is about 10$^{121}$ times larger than the observed value. To alleviate these problems, time varying dark energy models have been considered. For the various models of dynamical dark energy, one may refer the review\cite{Sami1} and the references there in. There are two main classes of dark energy models, the quintessence models\cite{Fujii,Carrol}, with equation of state $\omega>-1,$ and phantom dark energy models with $ \omega<-1. $ Compared to the quintessence form, phantom dark energy leads to unusual cosmological scenarios, like big-rip\cite{Caldwel1} where the universe may undergo super-exponential expansion, which effectively rip away the structures in the long run of the expansion of the universe.
 
There are attempts to explain the recent acceleration without invoking the exotic dark energy component. It was shown by several authors that a bulk viscous dark matter can cause an accelerated expansion of the universe. The effect of bulk viscosity was primarily analysed, in the context of acceleration in the early universe, the inflationary epoch\cite{Padmanabhan}. In the recent times, the effect of bulk viscous matter in causing the late acceleration 
was analyzed by many\cite{J. C. Fabris,B. Li,W.S. Hipolito,A. Avelino1,A. Avelino2,Athira,Athira2}. 

The detailed mechanism for the origin of bulk viscosity in the universe is still not correctly understood. From the theoretical point of view, the bulk viscosity can originate due to the deviation from the local thermodynamic equilibrium. It manifest as an effective pressure to bring back the system to its thermal equilibrium, which was broken when the cosmological fluid expands (or contracts) too fast. The bulk viscosity pressure thus generated, ceases as soon as the fluid reaches the equilibrium condition. 

There exists two main formalism to account for the bulk viscosity in cosmological theories, the non-causal theory where the dissipative perturbations propagate with infinite speed and causal theory, where the perturbations are propagating with finite speed. The non-causal formalism was developed by Eckart\cite{Eckart} and many used it in cosmology due to it's easiness in analysing the evolutionary behaviour of the cosmological parameters. Later Landau and Lifshitz\cite{Lifshitz} gives an equivalent formalism. The causal formalism was developed mainly by Israel, Stewart and Hiscock\cite{Israel1,Israel and J. M. Stewart,Israel2,Israel3,Hiscock}. 

In Eckart theory only the first order deviation from the equilibrium is considered, which effectively leads to the superluminal velocities of the dissipative signals, hence the theory is non-causal\cite{Israel1}. Moreover the resulting equilibrium states are unstable. However it illustrates a linear relationship between the bulk viscous pressure and the rate of expansion\cite{AColey2} of the universe. This facilitates the easy analytical method for the parameters in the context of expanding universe. 

Based on the Eckart formalism, Brevik and Gorbunova\cite{Brevik1} have shown that, the viscosity associated with matter, proportional to the expansion rate, can drive the universe into a phantom epoch. Fabris et al.\cite{J. C. Fabris} have considered a model with viscous coefficient proportional to $\rho^{\nu}$ (where $\rho$ is the density and $\nu$ is a constant) and have shown that, for $\nu=-\left(\alpha+\frac{1}{2}\right),$ ($\alpha$ is defined by the Chaplygin gas equation for pressre $p=-A/\rho^{\alpha}$) the model predicts a late acceleration similar to the generalized Chaplygin gas model of dark energy. They have also concluded that, even though the model is similar to the Chaplygin gas model at the background level, it does not show any oscillations in the power spectrum that plaugues the Generalized Chaplygin gas model. This can be considered as a positive indication of bulk viscous models. Later Avelino et al.\cite{A. Avelino3} have studied the bulk viscous matter model using Eckart formalism, where the bulk viscosity were taken to be proportional to both the velocity and acceleration of the universe. They have shown that the model can in general predicts the late acceleration of the universe. These authors did the asymptotic behaviour of this model also and argued that, the models is not stable assymptotically. Later in a more general anlysis on bulk visocus matter dominated model of the universe based on Eckart formalism by Athira and Mathew\cite{Athira}, have proved that the model have considerably good background evolution and asymptotically stable if the bulk viscous coefficient is a constant. All these analysis were based on the non-causal theory of viscosity. But for physically sound conclusions, one must use the causal theory of bulk viscosity. 

As mentioned earlier the causal theory of viscosity was proposed around 1979 by Israel and Stewart\cite{Israel2,Israel3}, which taking into account of the higher order deviations from the equilibrium, especially the second order deviations, which results the proper causal connection in the theory. Unlike Eckart theory, equilibria arises are stable. Moreover the Eckart theory can be obtained from it as a first order approximation. In certain studies, a truncated version of this theory has been used, where they omit some divergence terms in the expression for the evolution of the bulk viscous pressure, which contains terms corresponds to second order deviations\cite{HiscockLindblom,HiscockSalmonson} from equilibrium. 

Initial works where the full causal theory has been used are in the context of the inflation occured during the early period of the evolution of the universe. Maartens and Mendez\cite{R. Maartens} studied the early inflation caused by bulk viscous cosmic fluid using the full causal theory and found that the resulting solutions are thermodynamically consistent. Another interesting study on the cosmology of flat FLRW bulk viscous universe is in reference\cite{Mak and Harko}, where the authors find exact solutions corresponds to the early inflationary phase of the universe, with a bulk viscous coefficient proportional to the Hubble parameter. In a later work by the same authors using full causal theory, a new class of exact solutions were found by reducing the evolutionary equations of the universe to Abel-type first order differential equations\cite{MakHarko}. Zimdahl\cite{W.Zimdahl1} combined the equivalence between cosmological particle creation and effective viscous fluid pressure using Israel-Stewart model, and found that there exist an inherent self-limitation to the effective bulk viscous pressure due to the adiabatic particle production. They obtained solutions which indicates a transition from the inflationary phase to later non-inflationary epoch. Zakari and Jou\cite{M. Zakari} also studied the viscous driven inflationary epoch using the causal theory. Cooley et al.\cite{a a cooley} analysed the entropy production in a viscous universe using Israel-Stewart theory. Our concern here is the use of causal viscous formalism in analysing the late accelerating epoch of the universe. We would say that, such studies are comparatively less in number in the literature.

After the discovery of the late acceleration of the universe, the full causal theory of viscosity have been used to analyse the late stage of the universe having viscous cosmic components. Cataldo et al.\cite{M. Cataldo} have analysed the possibility of late acceleration, using Israel-Stewart formalism of bulk viscosity. In this work, the authors have used an ansatz for the Hubble parameter (inspired from the non-causal theory), and have shown that the universe might have undergone a transition to the phantom behaviour leading to big-rip singularity. Piattella et al.\cite{Piattella} have considered the bulk viscous universe using full causal theory with the aim of unifying dark matter and dark energy. They have found numerical solutions to the gravitational potential using an ansatz for the viscous pressure depends on the density of the viscous fluid and compared it with the standard $\Lambda$CDM model. Their conclusion is that, in gross comparison with the standard $\Lambda$CDM model, the viscous model with the full causal theory leads to some disfavoured features compared to the truncated version of the model. So by and large in solving the viscous model using the causal theory, it seems that, many have used some ansatz either for the Hubble parameter or for the viscous pressure. In the present study, we investigate the evolution of a bulk viscous matter dominated universe using the Israel-Stewart theory of bulk viscosity. We are trying to get the cosmic history by obtaining the analytical solutions of the Friedmann equations. 

The paper is organized as follows: in section \ref{FLRW universe dominated with bulk viscous matter} the Hubble parameter for the bulk viscous matter dominated universe is obtained, the behaviour of scale factor is analysed and the age of the universe is calculated. The evolution of cosmological parameters such as the deceleration parameter, the equation of state parameter, the matter density evolution and the curvature scalar have been discussed in section \ref{CosmologicalParameters}. In section \ref{thermodynamics} the validity of local and generalized second law of thermodynamics has been investigated. Section \ref{statefinder} and section \ref{parameterestimation} deals with the statefinder analysis and estimation of the model parameters respectively. The conclusions of this study is given section \ref{conclusion}
\section{FLRW universe dominated with bulk viscous matter}
\label{FLRW universe dominated with bulk viscous matter}
A spatially flat, homogeneous and isotropic universe is described by Friedmann-Lemaitre-Robertson-Walker (FLRW) metric,
\begin{equation}
ds^{2}=-dt^{2}+a(t)^{2}(dr^{2}+r^{2}d\theta^{2}+r^{2}sin^{2}\theta d\phi^{2}),
\end{equation}
where $(r,\theta,\phi)$ are the co-moving coordinates, $ t $ is the cosmic time and $ a(t) $ is the scale factor of the universe. The Friedmann equations describing the evolution of the flat universe, dominated with bulk viscous matter are
\begin{equation} \label{eqn:F1}
H^{2}=\frac{\rho_{m}}{3}
\end{equation}
\begin{equation}
2\frac{\ddot{a}}{a}+\left( \frac{\dot{a}}{a}\right)^{2}=-P_{eff}
\end{equation}
where $H=\frac{\dot{a}}{a}$ is the Hubble parameter, $ \rho_{m} $ is the matter density, $ P_{eff} $ is the effective pressure, an over dot represents the derivative with respect to cosmic time $t$ and we have taken $c= 8\pi G=1.$ The conservation equation for the viscous fluid is
\begin{equation} \label{eqn:con1}
\dot{\rho}_{m}+3H(\rho_{m}+P_{eff})=0.
\end{equation}
In these equations $P_{eff}$ is given as,
\begin{equation}
\label{eqn:effectivep}
P_{eff}=p+\Pi,
\end{equation}
where $ p $ is the normal pressure, given by $ p=(\gamma-1)\rho $, $ \gamma $ is the barotropic index and $ \Pi $ is the bulk viscous pressure. The radiation component is avoided and is a rational simplification as long as we are concerned with the late time evolution of the universe. The bulk viscous pressure in Eckart's theory is of the form $\Pi=-3H\xi$, where $ \xi $ is the term representing the bulk viscosity of the fluid and as a transport coefficient it can be a function of Hubble parameter of the universe. For sufficiently large $\xi $, it is possible that the negative pressure term can dominate and an accelerating cosmology can arise.

 According to Israel-Stewart causal theory, the effective pressure can satisfy the condition,
\begin{equation}\label{eqn:IS1}
\tau\dot\Pi+\Pi=-3\xi H-\frac{1}{2}\tau\Pi\left(3H+\frac{\dot\tau}{\tau}-\frac{\dot\xi}{\xi}-
\frac{\dot{T}}{T}\right),
\end{equation}
where $\tau$, $ \xi $ and $ T $ are the relaxation time, bulk viscosity and temperature respectively and are functions of the density of the fluid in general, defined by the following equations\cite{Maartens1}
\begin{equation}
\tau=\alpha\rho^{s-1}
\end{equation}
\begin{equation}
\label{bulk viscosity parameter equation}
\xi=\alpha\rho^{s}
\end{equation}
\begin{equation}
T=\beta\rho^{r}
\end{equation}
where $\alpha$, $\beta$ and $s$ are constant parameters satisfying the conditions, $ \alpha\geq0 $ and $ \beta\geq0 $ and $ r=\frac{\gamma-1}{\gamma} $. For relaxation time $\tau=0$, the differential equation for $\Pi$ reduces to the simple 
Eckart equation for the viscous pressure. Avoiding the second term on the right hand side of the equation will results in to the so called truncated equation.

\begin{figure}
\centering
\includegraphics[scale=1]{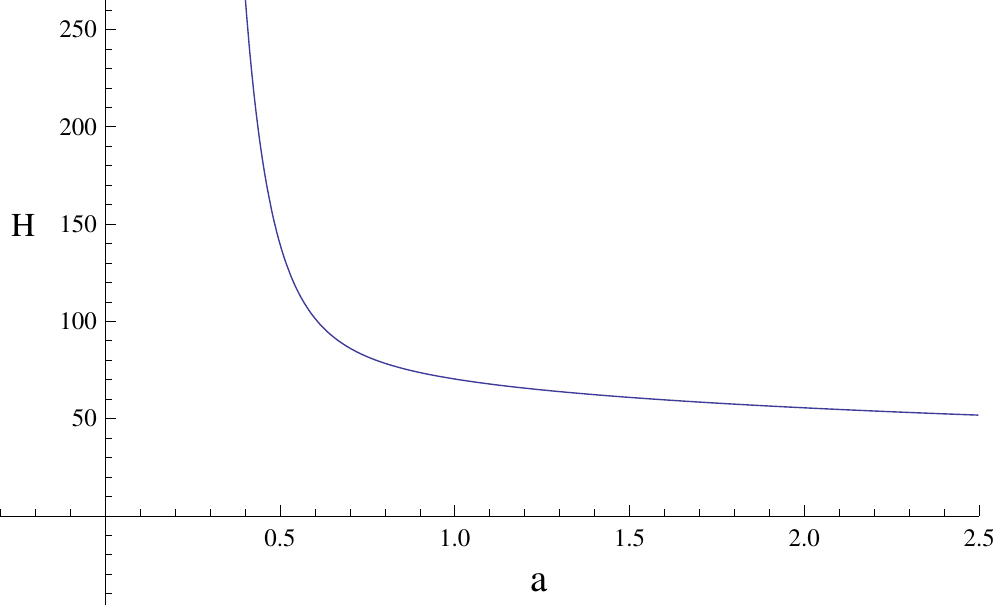}
\caption{The evolution Hubble parameter with scale factor for the best estimated values of the model parameters.}
\label{Hplot} 
\end{figure}

 Friedmann equation (\ref{eqn:F1}) can be combined with equations
 (\ref{eqn:con1}) and (\ref{eqn:effectivep}), to express the bulk viscous pressure $\Pi$ as
\begin{equation}
\Pi=-\left[2\dot{H}+3H^2+(\gamma-1)\rho\right],
\end{equation}
and the time derivative of it is,
\begin{equation}
\dot{\Pi}=-\left[2\ddot{H}+6H\dot{H}+(\gamma-1)\dot{\rho}\right].
\end{equation}
Then the evolution of the bulk viscosity as given by equation (\ref{eqn:IS1}), can be expressed as,
\begin{equation}
\begin{split}
\ddot{H} + \frac{3}{2}\left[1+(1-\gamma)\right]H\dot{H} + 3^{1-s}\alpha^{-1}H^{2-2s}\dot{H}-(1+r)H^{-1}\dot{H^{2}}+\\
\frac{9}{4}(\gamma-2)H^{3}+\frac{1}{2}3^{2-s}\alpha^{-1}\gamma H^{4-2s}=0,
\end{split}
\end{equation}
here we have used the density dependence of $\tau,$ $ \xi$ and $T$ as given previously.
We are considering non-relativistic matter, for which $ \gamma=1 $ and we also took $ s=\frac{1}{2} $\cite{ChimentoJacubi}, implies that the bulk viscosity is directly proportional to the Hubble parameter. The above equation then takes the form, 
\begin{equation}
\ddot{H}+b_1H\dot{H}-H^{-1}\dot{H^{2}}+b_2H^{3}=0,
\end{equation}
where $ b_1 $ and $b_2$ are taken as
\begin{equation}
b_1=3\left( 1+\frac{1}{\sqrt{3}\alpha} \right),  \, \, \, \, \, \, \, 
b_2=\frac{9}{4}\left( \frac{2}{\sqrt{3}\alpha}-1 \right).
\end{equation}
For calculational purpose we change the variable from cosmic time $t$ to $ x = \ln a $, the above differential equation become,
\begin{equation}
\frac{d^2 H}{d x^2}+b_1\frac{d H}{d x}+b_2H=0.
\end{equation}
On solving this we obtained the evolution of the Hubble parameter as,
\begin{equation}
\label{eqn:Hubbleparameter}
H=H_0\left(C_1 a^{-m_1}+C_2a^{-m_2}\right),
\end{equation}
where $ H_0$ is the present Hubble parameter,
\begin{equation}
\label{consatntC1}
C_1=\frac{1+\sqrt{1+6\alpha^{2}}-\sqrt{3}\alpha\tilde{\Pi_0}}{2\sqrt{1+6\alpha^{2}}},
\end{equation}
\begin{equation}
\label{consatntC2}
C_2=\frac{-1+\sqrt{1+6\alpha^{2}}+\sqrt{3}\alpha\tilde{\Pi_0}}{2\sqrt{1+6\alpha^{2}}},
\end{equation}
\begin{equation}
\label{constantm1}
m_1=\frac{\sqrt{3}}{2\alpha}\left(\sqrt{3}\alpha+1-\sqrt{1+6\alpha^{2}}\right)
\end{equation}
and
\begin{equation}
\label{constantm2}
m_2=\frac{\sqrt{3}}{2\alpha}\left(\sqrt{3}\alpha+1+\sqrt{1+6\alpha^{2}}\right).
\end{equation}
 In equations (\ref{consatntC1}) and (\ref{consatntC2}), $ \tilde{\Pi_0}=\frac{\Pi}{3H_{0}^{2}} $ is the dimensionless bulk viscous pressure parameter and constants $C_1$ and $C_2$ together satisfies, $ C_1 + C_2=1.$ However we could obtain the behaviour of Hubble parameter by numerical methods. For this we evaluated the parameters, $\alpha, \tilde{\Pi_0}$ and the present Hubble parameter $H_0$ in the present model using the cosmological data on supernovae (see section \ref{parameterestimation}). In Figure (\ref{Hplot}), the evolution of the Hubble parameter with scale factor corresponding to the best estimated values of model parameters is shown.
 In the limit of zero viscosity in the non-relativistic matter, implies $\alpha\to0$ equivalently viscous pressure $\Pi\to0,$ the Hubble parameter will reduces to,
$ H \sim a^{-{3}/{2}}$, which corresponds to the ordinary (non-viscous) matter dominated phase. 
Equation (\ref{eqn:Hubbleparameter}), also shows that, the Hubble parameter will becomes infinity as $a \to 0$. Hence the density will also becomes infinite at the origin, which suggests the presence of the big bang at the origin.
\subsection{Behaviour of scale factor}
\begin{figure}
\centering
\includegraphics[scale=1]{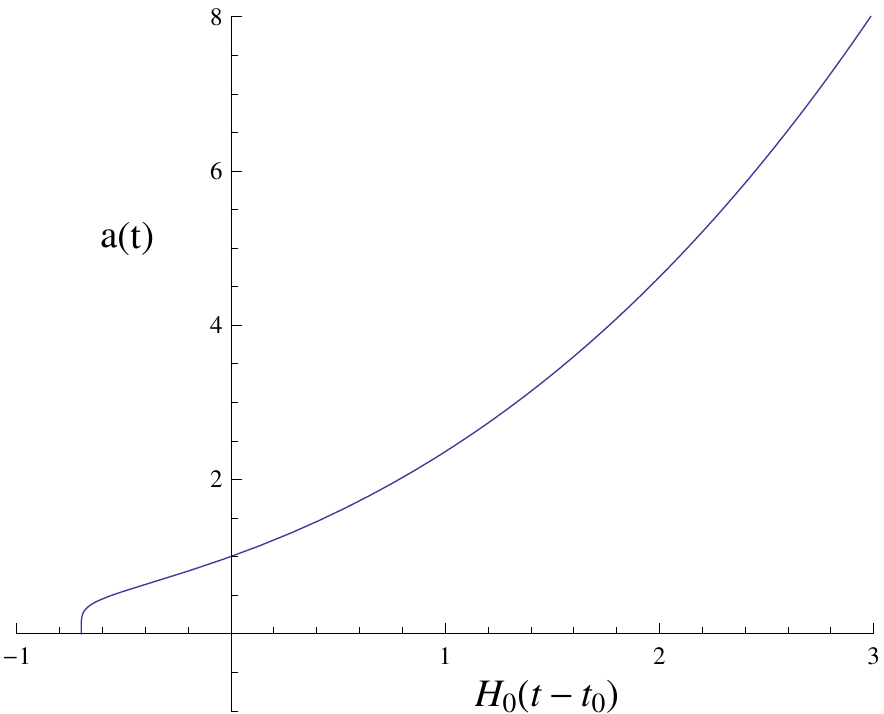}
\caption{The evolution of scale factor with $H_0(t-t_0)$ for the best estimated values of the model parameters.}
\label{scalefactor} 
\end{figure}

The Hubble parameter given in equation (\ref{eqn:Hubbleparameter}) can be integrated and is resulted into,
\begin{equation}
\label{scale factor}
\begin{split}
a^{m_1} {_2F_1}\left[1,\frac{m_1}{m_1-m_2},1+\frac{m_1}{m_1-m_2}\frac{a^{m_1-m_2} C_2}{C_1}\right]= C_1 m_1 H_0(t-t_0) + \\ {_2F_1}\left[1,\frac{m_1}{m_1-m_2},1+\frac{m_1}{m_1-m_2},-\frac{C_2}{C_1}\right],
\end{split},
\end{equation}
where ${_2F_1}[...]$ is the hyper-geometric function with respective arguments. It is to be noted that the hyper-geometric function on the left hand side itself depends on the scale factor. This equation can be used to assess the evolution of the scale factor. The nature of the evolution of $a(t)$ is not explicitly evident from the above equation, due to the appearance of the hyper-geometric functions in the equation. The behaviour of the scale factor with $ H_{0}(t-t_0) $ for the best estimated values of parameters is shown in Figure (\ref{scalefactor}).
 The scale factor is approximately linear at sufficiently early time, corresponds to the decelerated epoch and evolves exponentially with time in the extreme future epoch, corresponds to the de Sitter phase. This asymptotic behaviours indicates the transition from the early decelerated phase to a later accelerated expansion of the universe. The figure also shows that as $t \to 0$ the scale factor $a \to 0,$ indicating the presence of the big bang at the origin, hence age of the universe is properly defined. Compared to the corresponding model using Eckart formalism\cite{Athira} (in which the viscosity is taken to be proportional to both the velocity and acceleration of the universe) the evolution of the scale factor in the present model is almost similar. 

The transition redshift $z_T,$ corresponding to the switch over from deceleration to acceleration, can be obtained as follows. From the Hubble parameter in equation(\ref{eqn:Hubbleparameter}), the derivative of $ \dot{a} $ with respect to $ a $ can be written as,
\begin{equation}
\frac{d\dot{a}}{da}=H_{0}\left[C_{1}(1-m_{1})a^{-m_1}+C_2(1-m_{2})a^{-m_2}\right].
\end{equation}
Equating this to zero, we can get the transition scale factor $ a_{T} $ as,
\begin{equation}
\label{transition scale factor}
a_{T}=\left[-\dfrac{C_{2}(1-m_{2})}{C_{1}(1-m_{1})}\right]^{\frac{1}{m_{2}-m_{1}}},
\end{equation}
then the transition redshift $ z_{T} $ can be
\begin{equation}
\label{Transition redshift}
 z_{T}=\left[-\dfrac{C_{2}(1-m_{2})}{C_{1}(1-m_{1})}\right]^{-\frac{1}{m_{2}-m_{1}}}-1.
\end{equation}
For the best estimated values of the parameters, the transition red-shift, $ z_{T}\sim0.52^{+0.010}_{-0.016} $. This is within the WMAP range $z_T=(0.45-0.73)$ \cite{U. Alam}. In reference\cite{Athira}, the authors have estimated the transition redshift for a bulk viscous universe using Eckart formalism as around $0.49.$ In the present study using causal formalism, we have considered only a velocity dependence for the bulk viscous coefficient and the transition is found to be occurred slightly earlier. 

\subsection{The age of the universe}
\begin{figure}
\centering
\includegraphics[scale=1]{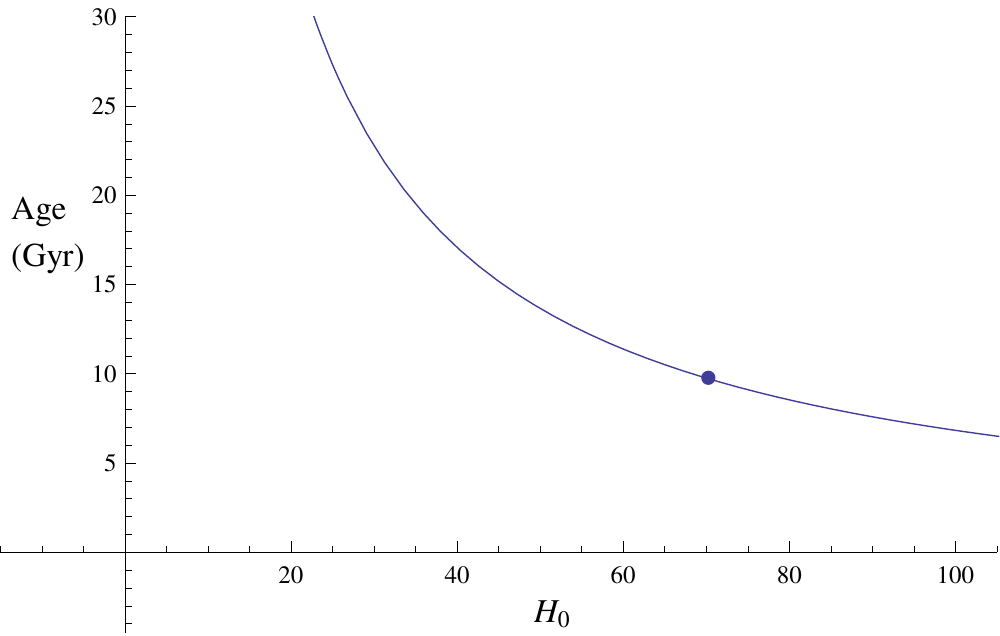}
\caption{The age of the universe in Gyr with $H_0$ in $km s^{-1} Mpc^{-1}$. The point marked in the plot corresponds to the age $9.72$ Gyr obtained in the model for the best estimated values of the parameters. }
\label{Agegraph}
\end{figure}

The age of the universe can be determined from the scale factor, equation(\ref{scale factor}).
On equating the scale factor to zero for $t=t_B,$ the big-bang time, the age $t_0 - t_B$ of the universe can be obtained. The age for different values of $H_0$ is as shown Figure (\ref{Agegraph}), where we have used the best estimate of the model parameters. An estimation using the scale factor equation will lead to a simple equation for age as,
\begin{equation}
t_0-t_B=0.6985 H_0^{-1}.
\end{equation}
The age of the universe corresponds to the best estimates of $ \alpha $, $ \tilde{\Pi_{0}} $ and $ H_0 $ is found to be around $9.72$ Gyr. This is considerably less than the standard value of age $13.74$ Gyr deduced from CMB anisotropy data\cite{M.Tegmark}, and $12.9 \pm 2.9$ Gyr from the oldest globular clustures\cite{E.Carretta}. Also the age from the present model is less than the age in the corresponding model using Eckart non-causal formalism, around $10.9 $ Gyr \cite{Athira}. 

\section{Evolution of other Cosmological parameters}
\label{CosmologicalParameters}
\subsection{The behaviour of deceleration parameter}
\label{q}
The deceleration parameter gives the measure of rate at which the expansion of the universe is taking place. If deceleration parameter is positive, then the universe is in decelerating phase and vice versa. The deceleration parameter $ q, $ can be expressed as
\begin{equation}\label{decelerationeqn}
q= -\frac{\ddot{a}a}{\dot{a}^{2}}=-\frac{\ddot{a}}{a}\frac{1}{H^{2}}=-1-\frac{\dot{H}}{H^{2}}.
\end{equation}
On substituting the Hubble parameter and it's derivative, the deceleration parameter become,
\begin{equation}
\label{eqn:q}
q(a)=-1 + \frac{C_{1} m_{1}a^{-m_{1}}+C_{2} m_{2}a^{-m_{2}}}{C_1 a^{-m_1}+C_2a^{-m_2}}.
\end{equation}
Using the best estimates of the model parameter, the coefficients takes the values, $m_1=0.31$ and $m_2=5.29.$ Hence as $a \to \infty$ the term $a^{-m_2}$ decreases faster than the term $a^{-m_1},$ hence $a^{-m_2}$ term can be neglected, consequently the deceleration parameter in this limit become, $q \to -1+m_1.$ On the other hand as $a \to 0$, the term $a^{-m_1}$ become negligibly small, as a result the deceleration parameter become, $q \to -1 + m_2.$
\begin{figure}
\centering
\includegraphics[scale=1]{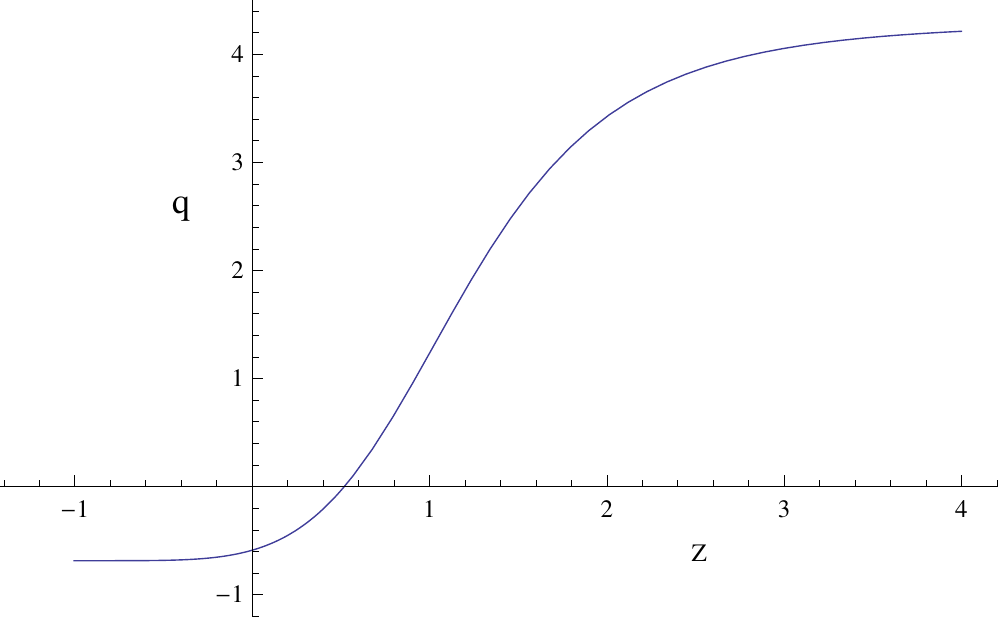}
\caption{ The evolution of deceleration parameter with respect to redshift for the best estimated values of the model parameters. }
\label{qfig}
\end{figure}
The deceleration parameter of the current epoch, corresponds to $ z=0 $ is, 
\begin{equation}
\label{present q}
q_{0}=-1+C_1{m_1}+C_2{m_2}=\frac{1}{2} \left(1+3 \tilde{\Pi _0}\right).
\end{equation}
For the best estimated values of $ \alpha  $ and $ \tilde{\Pi_0},$ the present value of the deceleration parameter is found to be $ q_0\sim-0.59^{+0.015}_{-0.016},$ which is quite near to the WMAP value, $q_0=-0.60$\cite{U. Alam}. By using Eckart theory and taking velocity and acceleration dependence for the the bulk viscous coefficient, the $q_0 \sim -0.64$ \cite{Athira}. The evolution of $q$ is as shown in Figure (\ref{qfig}). From the figure it is seen that, the deceleration parameter will be stabilizes around $-0.7$ in the far future of the evolution of the universe and is in confirmation with the previously obtained limit $q=-1+m_1 \sim -0.7.$ So even though the present model is predicting a never ending accelerating phase, the universe is not reaching the exact de Sitter phase and this is a marked deviation from the corresponding models using Eckart formalism\cite{Athira}, in which the model evolves asymptotically to the de Sitter phase. The evolution of $q$ parameter shows that the model is within the quintessence class. In the earlier epoch, the value of $q$ is positive and considerably large, around $q \sim 4,$ and is in confirmation with the previously obtained asymptotic limit $q=-1+m_2.$ While in the corresponding model with Eckart formalism, the deceleration parameter was found to be around $q \sim 2$ in the remote past of the universe\cite{Athira}. So the viscous matter behaves as a hard stiff fluid in the earlier epoch and evolves to a negative pressure fluid in the later phase. In this sense the present model is seems to be similar, except the fact that in the Eckart formalism the model will ultimately evolves to de Sitter phase, while in causal formalism the fluid comparatively more stiff in the earlier phase, but not ending with an exact de Sitter phase ultimately.

\subsection{Evolution of equation of state parameter}
The equation of state parameter $ \omega $ have a significant effect on the future expansion profile of the universe.   Universe enters the accelerating epoch when $ \omega<-\frac{1}{3} $. The equation of state can be obtained from the Hubble parameter using the relation,
\begin{figure}
\centering
\includegraphics[scale=1]{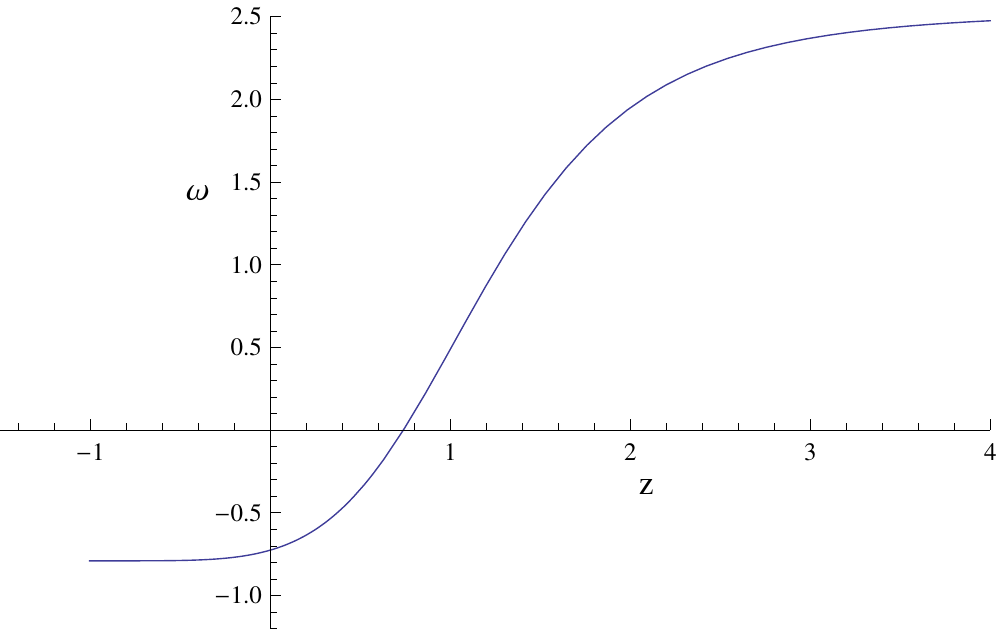}
\caption{The evolution of equation of state parameter with redshift for the best estimated value of model parameters.}
\label{equation of state figure}
\end{figure}
\begin{equation}
\omega=-1-\frac{1}{3}\frac{d \ln h^{2}}{dx},
\end{equation}
where $ h=\frac{H}{H_{0}} $ and $ x=\ln a $. 
Substituting the expression for the Hubble parameter, the equation of state is found to be,
\begin{equation}
\label{equation of state}
\omega=-1+\frac{2(C_1m_1 a^{-m_1}+C_2m_2a^{-m_2})}{3\left({C_1 a^{-m_1}+C_2a^{-m_2}}\right)}.
\end{equation}
Since as per the best estimates of the model parameters, $m_2>m_1,$ then the limiting condition $a \to \infty$ the equation of state $\omega \to -1+\frac{2}{3}m_1 \sim -0.79,$ which corresponds to quintessence nature. While at $a \to 0$ it becomes to $\omega \to -1+\frac{2}{3}m_2 \sim 2.5$ and is corresponds to stiff fluid characteristics. So the viscous matter evolution having a stiff fluid nature in past evolution to the characteristic of a fluid capable of providing negative pressure as the universe expands. The Eckart formalism approach, the bulk viscous matter behave as a stiff fluid with equation of state equal to +1, the equation of state become -1 corresponds to de Sitter phase.
The present value of equation of state parameter is
\begin{equation}
\omega_{0}=-1+\frac{2(C_{1}m_{1}+C_{2}m_{2})}{3},
\end{equation} 
and with the best estimated values of the model parameters, it comes around $\omega_0 \sim -0.73^{+0.01}_{-0.01}$ and is slightly higher than value obtained by the combined analysis of WMAP+BAO +$H_{0}$+SN data, around $-0.93$\cite{E.Komatsu,L.P.Chimanto}. The evolution of the equation of state parameter with redshift, for best estimated values of the model parameters is shown in Figure (\ref{equation of state figure}). At this juncture one should note the work by Brevik et al\cite{I.Brevik2} analysing the possibility of little rip in which the equation of state approach $-1$ asymptotically from below. In their work they have used a very special equation of state $p=-\rho-f(\rho)-\xi(H),$ where $f(\rho)$ is a chosen function of the density of bulk viscous matter and $\xi(H)$ is the bulk viscous pressure depends on the Hubble parameter. They have shown the possibility of a little rip with some assumed form for $f(\rho).$ On the other hand we haven't made any such pre-assumed form for equation of state. There are also some results that the bulk viscous matter can lead to phantom nature in the early period of the universe\cite{R. Maartens, Mak and Harko, MakHarko}. In the present analysis using the causal viscous formalism due to Israel and Stewart, we get into the result that the bulk viscous matter will behave like a strong stiff fluid in the early period and shows the behaviour of the quintessence dark energy in the later universe such that the equation of state stabilizes at around $\omega \sim -0.79$ in the far future of the evolution of the universe.
 \subsection{Evolution of the matter density}
The matter density parameter is defined by
\begin{equation}
\Omega_{m}=\frac{\rho_{m}}{\rho_{crit}},
\end{equation}
where $ \rho_{crit}=3H_{0}^{2} $ is the critical density. Using equations (\ref{eqn:F1}) and (\ref{eqn:Hubbleparameter}) we get,
\begin{equation}\label{Matterden}
\Omega_{m}(a)=(C_{1}a^{-m_{1}}+C_{2}a^{-m_{2}})^{2}.
\end{equation}
From the above equation the present matter density parameter $ \Omega_{m_{0}}$, can be obtained by taking the scale factor $ a=1 $. We get, 
\begin{equation}
 \Omega_{m_{0}}=(C_1+C_2)^{2}=1.
\end{equation}
Since in the present model, there is only matter as the major component. For zero bulk viscosity and bulk viscous pressure, the parameter takes the value $ C_{1}=1 $, $ C_{2}=0 $ and $ m_{1}\sim \frac{3}{2} $, correspondingly $ \Omega_{m}(a)\sim a^{-3} $, the usual matter dominated universe. From equation (\ref{Matterden}), it is seen that as $a \to \infty$ the density will go to $\Omega_m \to a^{-2m_1}$ while as $a \to 0$ it behaves as $\Omega_m \to a^{-2m_2}.$
\begin{figure} 
\centering
\includegraphics[scale=1]{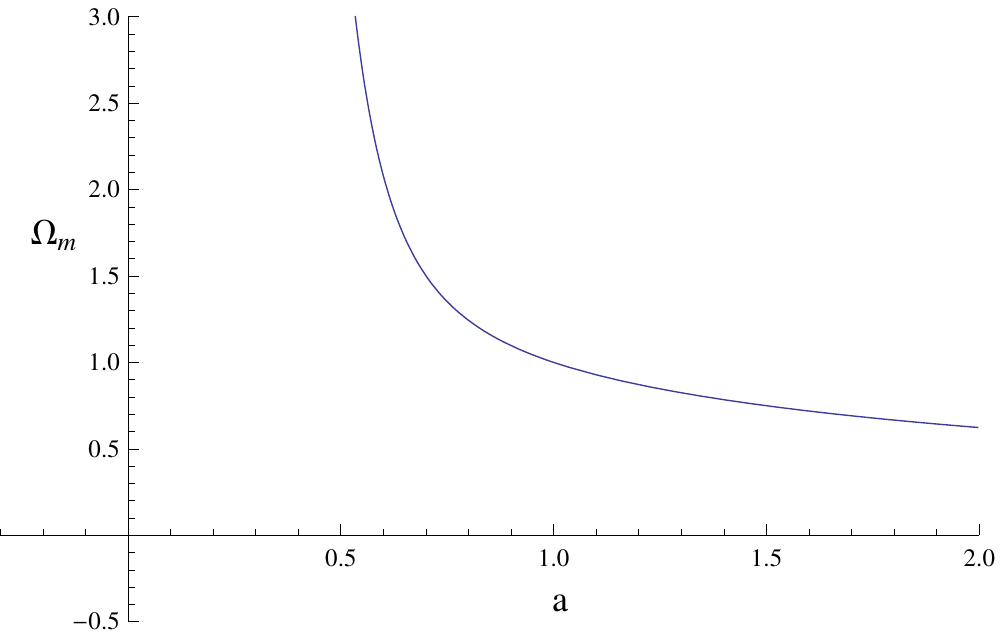}
\caption{The variation of matter density parameter with scale factor corresponding to the extracted values of the model parameters.}
\label{Evolution of matter density}
\end{figure}
 The evolution of matter density parameter with scale factor for best estimated values of $ \alpha $ and $ \tilde{\Pi_{0}}$ is shown in Figure (\ref{Evolution of matter density}). The figure shows that, as the scale factor $ a\rightarrow 0 $, the matter density increases rapidly and approach infinitely large value. This implies the big bang at the beginning of the universe. The decreasing nature of density in the future depicts the absence of the big-rip. In the overall way the evolution of density parameter in the present model is similar to that using Eckart formalism\cite{Athira}.

\subsection{Evolution of curvature scalar}
\begin{figure}
\centering
\includegraphics[scale=1]{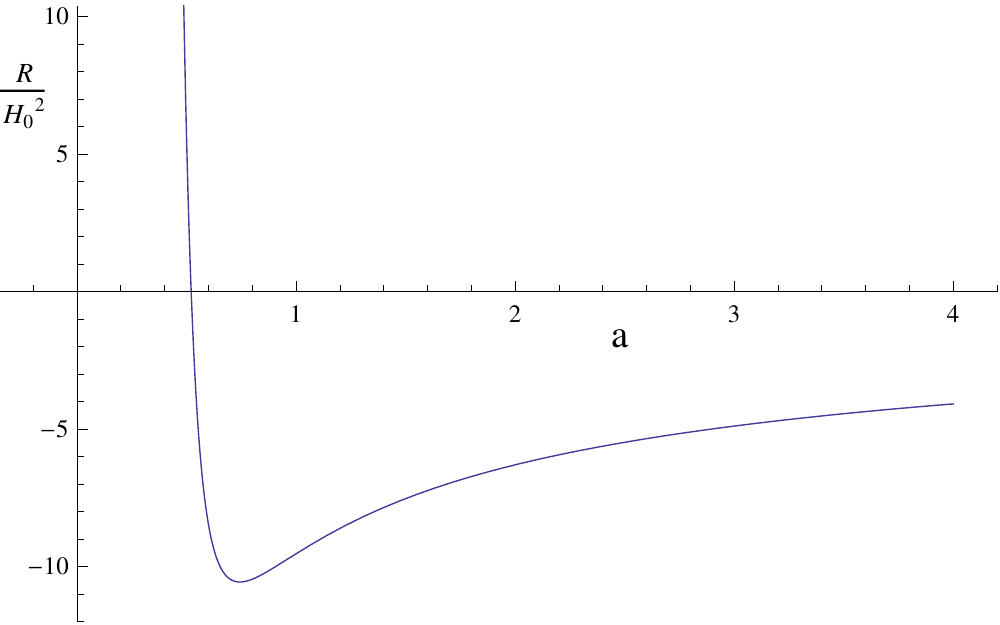}
\caption{The variation of curvature scalar against scale factor for the best estimated values of parameters.}
\label{R} 
\end{figure}
The evolution of curvature scalar of the universe enable one to confirm the occurrence of initial singularity in the model. The curvature scalar R, for a flat universe, is defined as\cite{Kolb}
\begin{equation}
R=-6\left(\dot{H}+2H^2\right).
\end{equation}
Using the Hubble parameter, evolution equation of the curvature scalar, can be obtained as
\begin{equation}
\label{curvaturesclar}
\begin{split}
R(a)=6 a^{-2 \left(m_1+m_2\right)} H_0^2 \left(a^{m_2} C_1+a^{m_1} C_2\right)\left[a^{m_2} C_1 \left(-2+m_1\right) +a^{m_1} C_2 \left(-2+m_2\right)\right].
\end{split}
\end{equation} 
The above equation shows that the curvature scalar $ R\to \infty$ when $ a \to 0 $ implies the initial singularity corresponding to the big bang. The behaviour of curvature scalar with scale factor for the best estimated values of model parameters is shown in Figure (\ref{R}). This evolution of the curvature scalar suggests presence of big bang at the origin of the universe. 
\section{Entropy and second law of thermodynamics}
\label{thermodynamics}
In the FLRW universe the bulk viscosity causes the production of local entropy. The law of production of local entropy on the FLRW space-time is expressed as\cite{S.Weinberg},
\begin{equation}
\label{local entropy}
T\nabla_{\nu}s^{\nu}=\xi(\nabla_{\nu}u^{\nu})=9H^{2}\xi,
\end{equation}
where $ T $ is the temperature and $ \nabla_{\nu}s^{\nu} $ is the rate of generation of entropy in unit volume. The condition for the validity of the second law of thermodynamics is then become,
\begin{equation}\label{eqn:2ndlaw}
T\nabla_{\nu}s^{\nu}\geq0.
\end{equation}
 This in turn implies that, the bulk viscosity must satisfies $ \xi\geq 0$.
From equation (\ref{bulk viscosity parameter equation}), the bulk viscosity for $s=1/2$ is,
\begin{equation}
\xi=\alpha\rho^{\frac{1}{2}}.
\end{equation}
 Substituting equations (\ref{eqn:F1}) and (\ref{eqn:Hubbleparameter}) in the above equation, we get
\begin{equation}
\label{bulk viscous parameter a}
\xi(a)=\sqrt{3} \alpha H_0(C_1 a^{-m_1}+C_2a^{-m_2}).
\end{equation}
\begin{figure}
\centering
\includegraphics[scale=1]{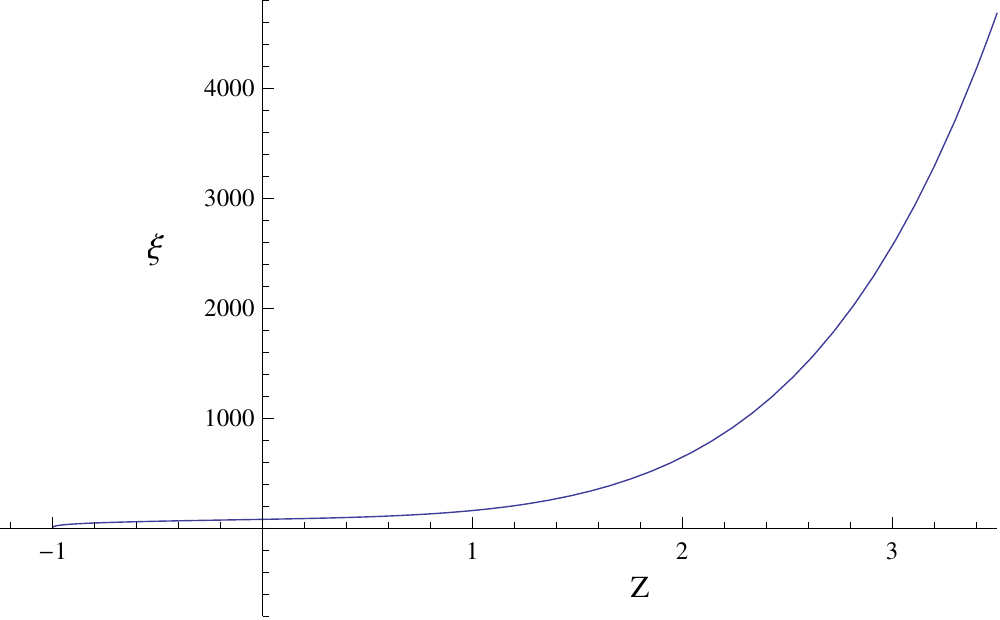}
\caption{The evolution of the bulk viscosity coefficient with redshift for the best estimated values of parameters.}
\label{evolution of bulk viscous parameter}
\end{figure}
For the best estimates of the model parameters $ m_2>m_1, $ the bulk viscosity has the following behaviour as $ a \to 0, $ the bulk viscosity satisfies, $ \xi \sim \sqrt{3} \alpha H_0 C_2 a^{-m_{2}} , $ since the magnitude of $m_2$ is large, $ \xi$ had large positive value in the early phase. While at $ a \to \infty $, it evolves to, $ \xi\sim \sqrt{3} \alpha H_0 C_1 a^{-m_{1}} $, here the value $ \xi$ will be small in future as the value $m_1$ is small and will be positive. Corresponding to the present epoch with $ a_0=1, $ the bulk viscosity become, $ \xi=\sqrt{3}\alpha H_{0}.$ All these together implies that the viscosity, $\xi >0$ always. This in turn implies that the condition given equation \ref{eqn:2ndlaw} always be satisfied, hence the second law of thermodynamics is satisified throughout the evolution of the universe. The evolution of $ \xi $ with respect to redshift z, for best estimated values of parameters is shown in Figure (\ref{evolution of bulk viscous parameter}). Therefore the rate of entropy production is always positive. Here also the present model shows remarkable difference from the Eckart formalism model. In analysing the model in Eckart formalism\cite{Athira}, it was found that the local second law of thermodynamics is violated during an early phase of the universe. Contrary to this there is no violation at all of the local second law in the present causal model. In this sense the causal model is to be favoured over the one based on Eckart formalism. Another distinct behaviour observed in the present model is that the bulk viscosity is decreases with time. Whereas the bulk viscosity coefficient is increasing from negative value region to a positive region in the model using Eckart formalism\cite{Athira}.

The entropy production from the horizon can also be accounted, which leads to the generalized second law (GSL), which states that the sum of total entropy of the fluid components of the universe and that of the horizon must always increase with time\cite{G.W.Gibbons,T.K.Mathew}. This can be expressed as,
\begin{equation}
\frac{d}{dt}(S_{m}+S_{h})\geq0,
\end{equation}
where $ S_{m} $ and $ S_{h} $ represents the entropy of matter and that of horizon respectively. The apparent horizon radius $ r_{A} $, for a spatially flat FLRW universe is given as\cite{A.Sheykhi}
\begin{equation}
\label{apparent horizon radius}
r_{A}=\frac{1}{H}.
\end{equation}
Using equations (\ref{eqn:F1}), (\ref{eqn:con1}) and (\ref{eqn:effectivep}) the time derivative of $ r_{A}  $ is obtained as,
\begin{equation}
\label{time derivative of apparent horizon radius}
\dot{r_{A}}=\frac{r_{A}^{2}}{2}(\Pi+\rho_{m}).
\end{equation}
 The entropy of the apparent horizon is proportional to the area of the Hubble horizon and is defined as\cite{P.C.W.Davis},
\begin{equation}
\label{entropy of horizon}
S_{h}=2\pi A=8\pi^{2} r_{A}^{2},
\end{equation}
where $ A=4\pi r_{A}^{2} $ is the area of the Hubble horizon. The time evolution of entropy of the horizon is
\begin{equation}
\label{time derivative of entropy of horizon}
\dot{S_{h}}=16\pi^{2}r_{A}\dot{r_{A}}.
\end{equation}
For the temperature of the apparent horizon we use the relation\cite{M.R.Setare}
\begin{equation}
\label{temperature horizon}
T_{h}=\frac{1}{2\pi r_{A}}\left(1-\frac{\dot{r_{A}}}{2 H r_{A}}\right).
\end{equation}
Using equations (\ref{apparent horizon radius}), (\ref{time derivative of apparent horizon radius}), (\ref{time derivative of entropy of horizon}) and (\ref{temperature horizon}), we can write
\begin{equation}
\label{change inentropy of horizon}
T_{h}\dot{S_{h}}=4\pi r_{A}^{2}(\Pi+\rho_{m})\left(1-\frac{\dot{r_{A}}}{2}\right).
\end{equation}
To determine the change in entropy of the matter component, we can apply the Gibb's relation,
\begin{equation}
T_{m}dS_{m}=dE+P_{eff}dV,
\end{equation}
where $ T_{m} $ is the temperature of the bulk viscous matter, $ E=\rho_{m}V, $ the total energy of the bulk viscous matter and $ V=\frac{4}{3}\pi r_{A}^{3} $ is the volume enclosed by the Hubble horizon. Using equation (\ref{eqn:effectivep}), the Gibb's equation becomes
\begin{equation}
\label{Gibb's equation}
T_{m}dS_{m}=Vd\rho_{m}+(\Pi+\rho_{m})dV.
\end{equation}
In thermal equilibrium, the temperatures of the viscous matter and horizon are equal, $ T_{m}=T_{h} $. The Gibb's equation (\ref{Gibb's equation}) can be re-written as,
\begin{equation}
\label{change in entropy of matter}
T_{h}\dot{S_{m}}=4\pi r_{A}^{2}(\Pi+\rho_{m})(\dot{r_{A}}-1).
\end{equation}
The time variation total entropy can be obtained by adding equations (\ref{change inentropy of horizon}) and (\ref{change in entropy of matter}),
\begin{equation}
\label{total entropy}
T_{h}(\dot{S_{h}}+\dot{S_{m}})=\frac{A}{4}r_{A}^{2}(\Pi+\rho_{m})^{2}.
\end{equation}
Since the radius and the area of the apparent horizon is always positive, from equation (\ref{total entropy}) it is evident that $ \dot{S_{h}}+\dot{S_{m}}\geq0 $, for a given temperature. Hence the GSL is satisfied. Therefore, the model in Eckart formalism\cite{Athira} and in the present model, the generalized second law of thermodynamics is satisfied. 

\section{Statefinder diagnostic}
In \cite{V.Sahni1} Sahni et al. have bring out a geometric diagnostic technique for contrasting various models of dark energy. Since all models predicting the Hubble parameter, scale factor, deceleration parameter etc., to distinguish
\begin{figure}
\centering
\includegraphics[scale=0.8]{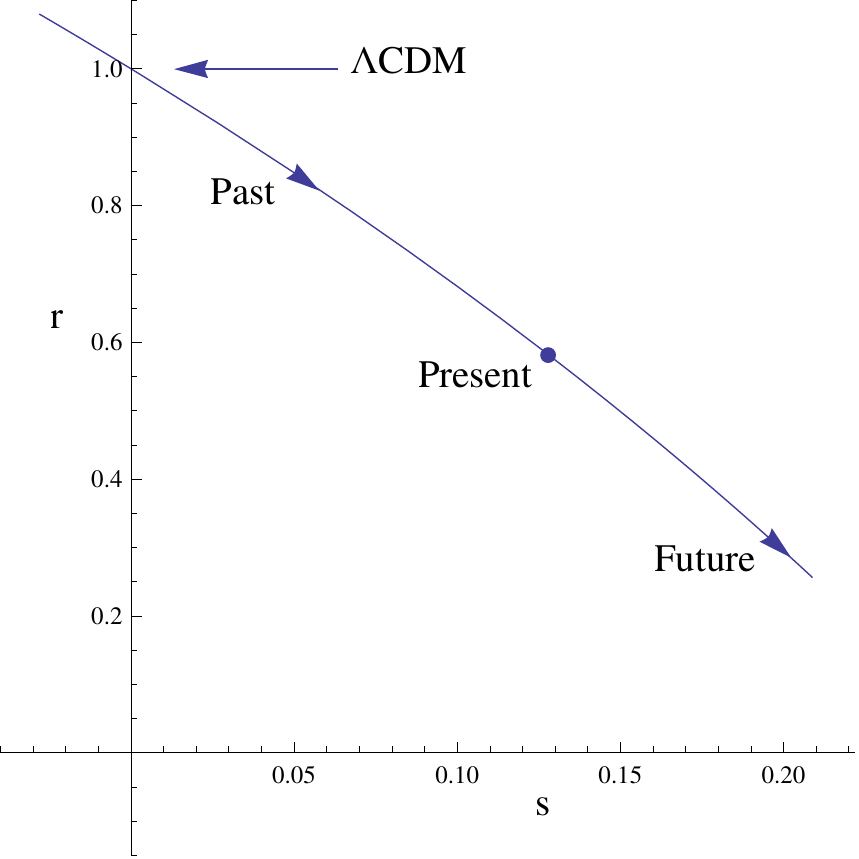}
\caption{The evolutionary trajectory of the bulk viscous model of the universe in the $r-s$ plane for the best estimated values of the model parameters.}
\label{Statefinderplot}
\end{figure} between the models, it is better to use quantities involving higher derivatives of $H$ or scale factor. The statefinder
 parameter pair $\lbrace r,s \rbrace$ introduced by them, depends on the the third order derivative of the scale factor. A characteristic property of statefinder parameter pair is that $ \lbrace{r,s}\rbrace=\lbrace{1,0}\rbrace,$ a fixed point for the $ \Lambda $CDM model. Evolutionary trajectories of these parameters for different dark energy models and their difference from the fixed $\Lambda$CDM point distinguishes the models from each other and also from the standard $ \Lambda $CDM model. The statefinder parameters are defined as 
\begin{equation}
\label{statefinderr}
r=\frac{\dddot{a}}{aH^{3}}=\frac{1}{2h}\frac{d^{2}h^{2}}{dx^{2}}+\frac{3}{2h^{2}}\frac{dh^{2}}{dx}+1,
\end{equation}
\label{statefinder}

\begin{equation}
\label{statefinders}
s=\frac{r-1}{3(q-\frac{1}{2})}=-\left(\frac{\frac{1}{2h}\frac{d^{2}h^{2}}{dx^{2}}+\frac{3}{2h^{2}}\frac{dh^{2}}{dx}}{\frac{3}{2h^{2}}\frac{dh^{2}}{dx}+\frac{9}{2}}\right).
\end{equation}
For the present bulk viscous matter dominated model, these parameters takes the form, 
\begin{equation}
\begin{split}
r=\frac{a^{2 m_2} C_1^2 \left(m_1-1\right) \left(2 m_1-1\right)+a^{2 m_1} C_2^2 \left(m_2-1\right) \left(2 m_2-1\right)}{\left(a^{m_2} C_1+a^{m_1} C_2\right){}^2}+\\
\frac{a^{\left(m_1+m_2\right)} C_1 C_2 \left(m_1+m_2-2\right) \left(m_1+m_2-1\right)}{\left(a^{m_2} C_1+a^{m_1} C_2\right){}^2}
\end{split}
\end{equation}
\begin{equation}
\begin{split}
s=\frac{2 \left(a^{2m_2} C_1^2 m_1 \left(2 m_1-3\right)+a^{2m_1} C_2^2 m_2 \left(2 m_2-3\right)\right)}{3 a^{2m_2} C_1^2 \left(2 m_1-3\right)+6 a^{\left(m_1+m_2\right)} C_1 C_2 \left(m_1+m_2-3\right)+3 a^{2m_1} C_2^2 \left(2 m_2-3\right)}+\\
\frac{2a^{\left(m_1+m_2\right)} C_1 C_2 \left(m_1+m_2-3\right) \left(m_1+m_2\right)}{3 a^{2m_2} C_1^2 \left(2 m_1-3\right)+6 a^{\left(m_1+m_2\right)} C_1 C_2 \left(m_1+m_2-3\right)+3 a^{2m_1} C_2^2 \left(2 m_2-3\right)}
\end{split}.
\end{equation}

The evolution of statefinder parameters in the r-s plane is shown in Figure (\ref{Statefinderplot}). The plot exhibits that the trajectory begins from the second quadrant of $ r-s $ plane, $ r>0 $ and $ s<0 $ and entered into the first quadrant, $\lbrace r,s\rbrace>0$. For the present universe, the statefinder parameters takes the form,
\begin{equation}
\label{presentr}
r_0=\frac{6 \sqrt{3} \tilde{\Pi _0}+\alpha  \left(22+9 \tilde{\Pi _0} \left(2+\tilde{\Pi _0}\right)\right)}{4 \alpha },
\end{equation}
\begin{equation}
\label{presents}
s_0=1+\frac{1}{\sqrt{3} \alpha }+\frac{1}{\tilde{\Pi _0}}+\frac{\tilde{\Pi _0}}{2}.
\end{equation}
Hence the present values of statefinder pair is $\lbrace r_0,s_0\rbrace=\lbrace 0.582,
0.128\rbrace$. This indicates that the present bulk viscous model is distinguishably different from the $ \Lambda $CDM model. From the Figure (\ref{Statefinderplot}) it can also infer that the bulk viscous model resembles the $\Lambda $CDM model during an early evolutionary phase of the universe and in the present time the model is moving away from $\Lambda $CDM model. In the first quadrant the trajectory is lying in the region $ r<1 $ and $ s>0 $, this represents the quintessence nature. This is a marked deviation from the corresponding model using Eckart formalism in which the model  approaches the $\Lambda$CDM model in the future\cite{Athira} in the $r-s$ plane.
\section{The model parameter estimation using Supernovae data}
\label{parameterestimation}
\label{Parameter estimation}
In this section we described the evaluation of the model parameters by constraining it with the observational data on type Ia supernovae. We have used ``Union" SNe Ia data set\cite{M.Kowalski}, consists of 307 Supernovae type Ia from 13
\begin{figure}[h]
\centering
\includegraphics[scale=1]{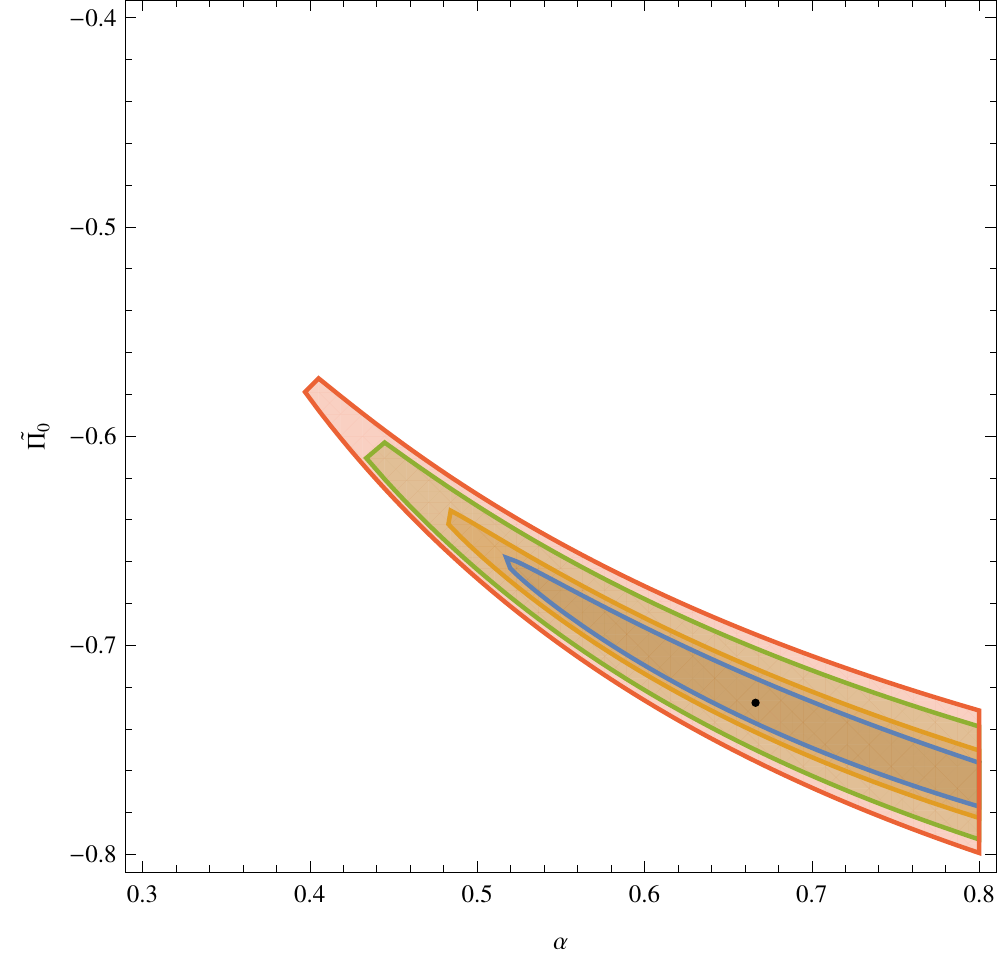}
\caption{The confidence intervals for the model parameters $\alpha$ and $\tilde{\Pi_0}$ correspond to $68.3\%,$ $95.4\%,$ $99.73\%,$ and $99.99\%$ probabilities. The best estimated values of the model parameters is indicated by the point. }
\label{Confidenceinterval}
\end{figure} 
 independent data sets. Our aim here is to extract the best fit for the parameters $ \alpha $ , $ \tilde{\Pi}_{0} $ and the present value Hubble parameter $ H_{0}. $ We obtained parameter values by applying the $ \chi^{2} $ minimization method. 

The luminosity distance $ d_{L} $ in flat universe is,
\begin{equation}
d_{L}(z,\alpha,\tilde{\Pi_{0}},H_{0})=c(1+z)\int_{0}^{z} \frac{dz'}{H(z',\alpha,\tilde{\Pi_{0}},H_{0})}.
\end{equation}
The difference between apparent and absolute magnitudes of supernovae is depends on the distance. The equation that
 relates the theoretical distance moduli $ \mu_{th} $, apparent magnitude $ m $, absolute magnitude $ M $ and $ d_{L} $ is given by
\begin{equation}
\label{distancemodulii}
\mu_{th}(z,\alpha,\tilde{\Pi_{0}},H_{0})=m-M=5\log_{10}\left[\frac{d_{L}(z,\alpha,\tilde{\Pi_{0}},H_{0})}{Mpc}\right]+25.
\end{equation}
The observational distance modulus $ \mu'_{i} $, obtained from SNe Ia data set is compared with $ \mu_{th} $ calculated using equation (\ref{distancemodulii}) corresponding to different values of redshifts. The $ \chi^2 $ function can be written as
\begin{equation}
\chi^{2}(\alpha,\tilde{\Pi_{0}},H_{0})=
‎\sum_{i=1}^n ‎\frac{\left[\mu_{th}(z,\alpha,\tilde{\Pi_{0}},H_{0})-\mu'_{i}\right]^2}{\sigma_{i}^2},
\end{equation}
where $ n $ is the total number of data points and $ \sigma_{i}^2 $ is the variance of the $i^{th}$ measurement. The best
\begin{table}
 \centering
\begin{tabular}{ |p{2cm}|p{1.7cm}|p{1.9cm}|p{0.8cm}|p{2.4cm}|p{1cm}|p{0.8cm}| }
 \hline
 \vspace{0.1cm}&\multicolumn{6}{|c|}{Estimated values of parameters} \\[8pt]
 \hline
 \vspace{0.1cm}Model&\vspace{0.1cm}$ \alpha $&\vspace{0.1cm}$  \tilde{\Pi_{0}}$&\vspace{0.1cm}$ \Omega_{m0} $&\vspace{0.1cm}$ H_{0} $\begin{small}
 ($ kms^{-1}Mpc^{-1} $)
 \end{small}&\vspace{0.1cm}$ \chi^{2}_{min} $&\vspace{0.1cm}$ \chi^{2}_{d.o.f.} $\\ [8pt] 
\hline
\vspace{0.1cm}
Bulk viscous model &\vspace{0.1cm} $0.665^{+0.030}_{-0.025}$ &\vspace{0.1cm}$-0.726^{+0.01}_{-0.01}$ &\vspace{0.1cm}1& \vspace{0.1cm}70.29& \vspace{0.1cm}310.29&\vspace{0.1cm} 1.020\\[8pt]
\hline 
\vspace{0.07cm}$\Lambda$CDM model&\vspace{0.07cm}-&\vspace{0.07cm}-&\vspace{0.07cm}0.316&\vspace{0.07cm}70.03& \vspace{0.07cm} 311.93 &\vspace{0.07cm}1.026\\[8pt]
\hline
\end{tabular}
\caption{A comparison of best estimated values of bulk viscous model parameters with the standard $\Lambda$CDM model. }
\label{Table:1} 
\end{table}
 estimate values of the parameters $ \alpha,\tilde{\Pi_{0}}$ and $H_{0} $ has been obtained by $ \chi^{2} $ minimization.

  The parameter values for the $ \Lambda $CDM model have also been extracted using the same data set for comparison. The best estimated parameter values is shown in the Table (\ref{Table:1}). The $ \chi^{2}_{min} $ function per degrees of freedom is defined as,
 $ \chi^{2}_{d.o.f.} =\frac{\chi^{2}_{min}}{n-n'}$, $ n' $ is the number of parameters in the model. Here $ n=307 $ and $ n'=3 $. The present value of the Hubble parameter obtained from the bulk viscous model is comparable with that of the $ \Lambda $CDM model.

 The confidence interval plane for the model parameters $ \alpha $ and $\tilde{\Pi_{0}} $ is shown in Figure (\ref{Confidenceinterval}). The contours are corresponding to $68.3\%, 95.4\%,
 99.73\%$ and $99.99\%$ probabilities as one move from inside. The probabilistic correction to the parameters value corresponds to 68.3\% probability have been shown in the table. It is these values which have been used to generate the evolutionary status of various cosmological parameters present in the previous sections.
\section{Conclusion}
\label{conclusion}
In the present work, the evolution of flat FLRW bulk viscous non-relativistic matter dominated universe has been investigated. We have used relativistic second order full causal theory for the evolution of the bulk viscous pressure. The evolution of the viscous pressure is as given by the differential equation (\ref{eqn:IS1}). The bulk viscous coefficient is taken as $\xi=\alpha \rho^{1/2},$ means it is effectively depending on the expansion velocity of the universe. We solved the Friedmann equations analytically to obtain the Hubble parameter as given equation (\ref{eqn:Hubbleparameter}). In the limit of zero viscosity, the present model reduces to the non-viscous matter dominated universe satisfying $ H \sim a^{-\frac{3}{2}}.$ 

We obtained the scale factor of the expansion also. The asymptotic behaviour of scale factor indicates the transition from an early decelerated to a late accelerated epoch and it's evolution as shown Figure (\ref{scalefactor}), indicates the presence of big bang at the origin of the universe. Hence the age of the universe is defined and is determined using the best estimated values of the parameters, is around $9.72$ Gyr. This is considerably less than the age obtained from the CMB anisotropic data and from the data of oldest globular clusters.

 The evolution of the deceleration parameter $ q, $ is obtained, as shown in Figure (\ref{qfig}). From this the transition red-shift was obtained, $z_{T}\sim0.52^{+0.010}_{-0.016}.$ The present value of the deceleration parameter is obtained as $ q_0\sim-0.59^{+0.015}_{-0.016},$ and is in the range obtained by WMAP data analysis. As $ a \to \infty $, $q$ will be stabilizes around the value $ -0.7. $ So unlike in the Eckart formalism approach, where the bulk viscous universe ultimately go over to a de Sitter phase, the present model is lying well within the range of quintessence behaviour asymptotically. In it's over all evolution, the deceleration parameter begins with $ q \sim 4 $ in the remote past and stabilises around $-0.7$ in the far future of the universe. 

The evolution of equation of state parameter $\omega$, in this model is obtained in equation (\ref{equation of state}) and the variation of it with redshift is shown in Figure (\ref{equation of state figure}). The present value of the equation of state parameter is obtained as $\omega_0 \sim -0.73^{+0.01}_{-0.01}.$ Hence the present universe is accelerating and acceleration has been begun in the recent past. The value of $\omega_0$ obtained in this model is slightly higher than that obtained from WMAP+BAO+$H_{0}$+SN data. The future evolution of $\omega$ indicates a never ending acceleration phase but not approaching the de Sitter phase. This is a marked deviation from the corresponding model using Eckart formalism, in which the expansion ultimately ends up with a de Sitter epoch. From the evolution of $\omega$ it was found that the equation of state with $ +2.5 $ in the remote past and evolves to $-0.79$ in the future stages. This indicates that the bulk viscous matter shows a stiff fluid characteristics in the earlier epoch and then evolves to the quintessence nature in the later stages. 

The matter density parameter $ \Omega_m, $ obtained in the present model is given in equation (\ref{Matterden}) and it's evolution with respect to scale factor is shown in Figure (\ref{Evolution of matter density}). The evolution of the matter density as $a \to 0,$ indicates the presence of the big bang at the origin of the universe. The decrease in density along with the expansion of the universe suggests the absence of big-rip in the future. For zero bulk viscosity the the density reduces to $\Omega \sim a^{-3}$ corresponds to the ordinary matter dominated era. In the overall way the behaviour of $\Omega_m$ in the present model is similar to that using Eckart formalism\cite{Athira}.

The evolution of curvature scalar $R,$ is given in equation (\ref{curvaturesclar}) and is plotted with scale factor in Figure (\ref{R}). When $a \to 0, $ $R \to \infty$ implying the presence of the big bang at the origin of the universe. 

The evolution of the bulk viscosity in the present model shows that it starts with large positive value in the early phase of the evolution and evolves to smaller values during the later evolutionary phase of the universe. Throughout the evolution it satisfies the condition $\xi\geq0$. Therefore, the entropy production is always positive and hence, the local second law of thermodynamics is satisfied in the present model. Here also the present model differs from the model using Eckart theory\cite{Athira}. In a model using Eckart formalism, the coefficient of bulk viscosity is increases from negative to positive values as the universe expands, hence the local second law is violated in the early epoch. The generalized second law is valid in the present model as like in the non-causal Eckart model of the bulk viscous universe.

In order to contrast the present model from the standard dark energy models the statefinder geometric diagnostic has been carried out. The evolution of the model in the $r-s$ plane is as shown in Figure (\ref{statefinder}). The present value of statefinder parameter pair is obtained as, $\lbrace r_0,s_0\rbrace=\lbrace 0.582,0.128\rbrace$, this suggests that the present model is distinct from the standard $\Lambda$CDM model. The present bulk viscous model resembles the $\Lambda$CDM model in an early phase of evolution. The evolution of the trajectory on the $r-s$ plane is lying in the region $r<1$ and $s>0$, this indicates the quintessence nature of bulk viscous matter in the later stages of the expanding universe. In contrast, in the model using Eckart formalism, the model approaches the $\Lambda$CDM model in the future and is a marked difference from the causal approach.

\end{document}